\documentclass[12pt]{iopart}

\usepackage{iopams}
\usepackage{float}
\usepackage{amssymb}
\usepackage{setstack}
\usepackage{graphicx}
\usepackage{dcolumn}   
\usepackage{bm}
\usepackage{dsfont}
\usepackage{amssymb}
\usepackage{subfigure}

\newcommand{\bvphi}{\mbox{\boldmath$\varphi$}}

\newcommand{\bC     }{\mbox{\boldmath$C$}}

\newcommand{\bK     }{\mbox{\boldmath$K$}}

\newcommand{\bH     }{\mbox{\boldmath$H$}}
\newcommand{\bT    }{\mbox{\boldmath$T$}}
\newcommand{\bI     }{\mbox{\boldmath$I$}}

\newcommand{\bA     }{\mbox{\boldmath$A$}}

\newcommand{\bZ     }{\mbox{\boldmath$Z$}}

\newcommand{\bM     }{\mbox{\boldmath$M$}}

\begin{document}
\date{\today}
\title[Eigenvalue fluctuations of Gaussian random matrices]
{\bf 
Replica-symmetric approach
to the typical eigenvalue fluctuations of Gaussian random matrices
}
\author{Fernando L. Metz}
\address{Instituto de F\'isica, Universidade Federal do Rio Grande do Sul, 91501-970 Porto Alegre, Brazil \\
  Departamento de F\'isica, Universidade Federal de Santa Maria, 97105-900 Santa Maria, Brazil \\
  London Mathematical Laboratory, 14 Buckingham Street, London WC2N 6DF, United Kingdom
}
\begin{abstract}
We discuss an approach to compute the first and second moments of the number of eigenvalues $I_N$ that lie in an arbitrary interval of 
the real line for $N \times N$ Gaussian random matrices. The method combines the standard replica-symmetric theory with a perturbative expansion
of the saddle-point action up to $O(1/N)$ ($N \gg 1$), leading to the correct logarithmic scaling of 
the variance $\langle I_{N}^{2} \rangle - \langle I_N \rangle^2 = O(\ln N)$ as well as to an
analytical expression for the $O(1/N)$ correction to the  average  $\langle I_N \rangle/N$. 
 Standard results for the
number variance at the local scaling regime are recovered in the limit of a vanishing interval.
The limitations of the replica-symmetric method are unveiled by comparing our results with those derived through
exact methods.
The present work represents an important step to study the fluctuations of $I_N$ in non-invariant random matrix ensembles, where 
the joint distribution of eigenvalues is not known. 
\end{abstract}


\section{Introduction}

In the last decades, random matrices have established itself as a fundamental tool
to model complex disordered systems, with many applications in different branches of science \cite{BookMehta,oxford}.
The primary aim in random matrix theory is the study of the eigenvalue statistics, which
is accessed by computing suitably chosen observables. 
Important observables are the moments of the number of eigenvalues $I(a,b)$ that
lie between $a \in \mathbb{R}$ and $b \in \mathbb{R}$. For $a \rightarrow -\infty$, the random variable $I$ is the
so-called index, i.e., the number of eigenvalues contained in the unbounded interval $(-\infty,b]$.
A great deal of attention has been devoted to the index of random matrices, especially due to its
ability to probe the stability properties of the energy landscape characterising disordered systems \cite{Cavagna2000,wales2003energy}.
Another prominent 
observable, derived from $I(a,b)$ and studied originally by Dyson \cite{Dyson2}, is the variance of the number of eigenvalues within the
bounded interval $[-L,L]$ ($L > 0$), also known as the number variance. Recently, this observable has been applied
to study the statistics of non-interacting fermions confined in a harmonic trap \cite{Marino2014,Marino2016}, and
the ergodicity of the eigenfunctions in a mean-field model for the Anderson localisation transition \cite{MetzLett,Metz2017}.

A powerful tool to study the fluctuations of $I(a,b)$ is the Coulomb gas approach \cite{Dyson1}, whose
starting point is an exact correspondence between the joint distribution of eigenvalues and the partition
function of a two-dimensional Coulomb gas confined to a line. The Coulomb gas technique has
been employed to derive analytical results for typical and atypical fluctuations of $I(a,b)$ in the case of rotationally 
invariant random matrices (RIRM), including Gaussian \cite{Marino2014,Marino2016,Majumdar1,Majumdar2,Perez2014b,Isaac2016}, Wishart 
\cite{Vivo,Perez2015} and Cauchy \cite{Majumdar3} random matrices.
Restricting ourselves to typical fluctuations of $I$, one of the central outcomes for RIRM 
is the logarithmic increase of the variance $\langle I^{2} \rangle - \langle  I \rangle^2 \propto \ln N$  as a function
of the matrix dimension $N \gg 1$, due to the strong correlations among the eigenvalues.
In spite of the success of the Coulomb gas method, its application is strictly limited to 
RIRM, where the joint distribution of eigenvalues is analytically known and, consequently, the analogy with the Coulomb gas partition 
function can be readily established.

Recently, a novel technique to study the fluctuations of $I(a,b)$ has been developed \cite{MetzLett,Metz2015}. This method
is based on the replica-symmetric theory of disordered systems and its main advantage lies in its broad range
of possible applications, which goes beyond the realm of RIRM. In fact, the replica approach allows
to derive analytical results for typical and atypical fluctuations of $I(a,b)$ in the case of random matrices where the joint distribution of eigenvalues
is not even known. The most typical example in this sense is the adjacency matrix
of sparse random graphs \cite{MetzLett,Metz2015}, where the eigenvalues behave as uncorrelated random 
variables and the leading term of the variance scales as $\langle I^{2} \rangle - \langle  I \rangle^2 \propto N$ for large $N$.
However, it is unclear whether the formalism of \cite{MetzLett,Metz2015} is able to 
grasp the variance behaviour $\langle I^{2} \rangle - \langle  I \rangle^2 \propto \ln N$ of random matrices
with strong correlated eigenvalues. In the case of a positive answer, this would pave the way
to inspect the fluctuations of $I$ in random matrices that are not rotationally invariant, but in which
the eigenvalues are strongly correlated. The ensembles of Gaussian random matrices are the
ideal testing ground for this matter, since one can make detailed comparisons with exact results
and identify eventual limitations of the replica method.

In this work we show how the replica approach, as developed in  \cite{Metz2015,MetzLett}, has to be further
adapted in order to derive the correct logarithmic scaling  $\langle I^{2} \rangle - \langle  I \rangle^2 \propto \ln N$
for the GOE and the GUE ensembles of random matrices \cite{Dyson1,BookMehta}.
Following the approach of \cite{MetzPar}, the central idea here consists in writing the characteristic function of $I$ as a saddle-point
integral, in which the action is computed perturbatively around its $N \rightarrow \infty$ limit.
We show that the leading term $\langle I^{2} \rangle - \langle  I \rangle^2 \propto \ln N$ ($N \gg 1$) 
is correctly recovered only when the fluctuations due to finite $N$ are taken into account. However, the 
present approach does not yield the exact expression for the $O(N^0)$ contribution to the variance of $I$, due 
to our assumption of replica-symmetry for the order-parameter.
In the case of the number variance,
our analytical expression converges, in the limit $L \rightarrow 0^+$, to standard results of
random matrix theory \cite{BookMehta}, valid in the regime where the spectral window is measured in units of the mean level spacing.
This result supports the central claim of reference \cite{Metz2017}, where the limit $L \rightarrow 0^+$ of the number variance is employed
to study the ergodic nature of the eigenfunctions in the Anderson model on a regular random graph.
As a by-product, our method yields the $O(1/N)$ correction to the intensive average $\langle I \rangle/N$, whose
exactness is tested against numerical diagonalisation and previous analytical results. While
the $O(1/N)$ contribution to $\langle I \rangle/N$ is exact for the GOE ensemble, it fails in the case
of the GUE ensemble due to our assumption of replica symmetry.
The present approach can
be also employed to compute the moments of characteristic polynomials of non-invariant random matrices, where
the key quantity under study is similar to eq. (7). The moments of characteristic polynomials
of invariant random matrices have been largely studied \cite{Brezin,Brezin1,Borodin}, in view of their connection
with the distribution of zeros of the Riemann zeta function \cite{keating}.

The paper is organised as follows. In the next section we show how the computation of the characteristic function
of $I$ can be pursued using the replica approach. Section \ref{bla1} explains the basic steps of the
replica method, including the perturbative calculation of the action up to $O(1/N)$. We 
make the replica-symmetric
assumption for the order-parameter and derive an analytical expression for the characteristic function in section \ref{bla2}.
Section \ref{bla3} derives the analytical results for the index, the number variance and the fluctuations
of $I$ in an arbitrary interval. The last section summarises our work and discusses the impact of replica symmetry in our
results. 

\section{The number of eigenvalues inside an interval} \label{bla}

We study here two different ensembles of $N \times N$ Gaussian random matrices $\bM$ with real eigenvalues $\lambda_1,\dots,\lambda_N$, defined
through the probability distribution $\mathcal{P}(\bM)$. If the elements of the ensemble are real symmetric matrices, we have
the Gaussian orthogonal ensemble (GOE) \cite{BookMehta}
\begin{equation}
\mathcal{P}(\bM) = \mathcal{N}_{GOE} \exp{\left(  - \frac{N}{4}  \Tr \bM^2 \right)}\,.
\label{goe}
\end{equation}
If the ensemble is composed of complex-Hermitian random matrices, we have the Gaussian unitary ensemble (GUE) \cite{BookMehta}
\begin{equation}
\mathcal{P}(\bM) = \mathcal{N}_{GUE} \exp{\left[  - \frac{N}{2}  \Tr \left( \bM^{\dagger} \bM \right) \right]}\,,
\label{gue}
\end{equation}
where $(\dots)^{\dagger}$ represents the conjugate transpose of a matrix. The explicit form of the normalization
factors in eqs. (\ref{goe}) and (\ref{gue}) are not important in our computation.

The number of eigenvalues $I_N(a,b)$ lying between $a \in \mathbb{R}$ and $b \in \mathbb{R}$ is given by
\begin{equation}
I_N(a,b) =  \sum_{\alpha=1}^N \left[ \Theta(b-\lambda_{\alpha}) -  \Theta(a-\lambda_{\alpha})   \right] \quad a < b,
\end{equation}
with $\Theta(\dots)$ the Heaviside step function. The statistics of $I_N(a,b)$ is encoded in the characteristic function 
\begin{equation}
\mathcal{G}_N (\mu) = \langle  \exp{\left[ i \mu  I_N(a,b)  \right]}  \rangle,
\label{ssqw}
\end{equation}
where $\langle \dots  \rangle$ is the ensemble average over the random matrix elements. In
particular, the first two moments of $I_N(a,b)$ read
\begin{equation}
\langle I_N(a,b) \rangle = - i \frac{\partial \mathcal{G}_N (\mu) }{ \partial \mu} \Big{|}_{\mu=0} ,
\qquad
\langle I_N^{2}(a,b) \rangle = -  \frac{\partial^2 \mathcal{G}_N (\mu) }{ \partial \mu^2} \Big{|}_{\mu=0} .
\end{equation}

In order to calculate the ensemble average in eq. (\ref{ssqw}), one has to write $I_N(a,b)$ in terms
of the random matrix $\bM$. By following \cite{Cavagna2000,Metz2015} and representing $\Theta$ as the discontinuity 
of the principal value of the complex logarithm along the negative real axis, $I_N(a,b)$ may be written as the limit
\begin{equation}
\fl
I_N(a,b) = \frac{1}{2 \pi i} \lim_{\eta \rightarrow 0^{+}} \ln{\left[ \frac{Z(z_b) Z(z^{*}_a) }{Z(z^{*}_b)Z(z_a)  }    \right]} ,
\qquad z_a = a + i \eta \, , \qquad z_b = b + i \eta \, ,
\label{eod}
\end{equation}
with $Z(z) = \left[ \det{\left( \bM -z \bI_N  \right)}  \right]^{-1}$. Here $z$ is an arbitrary complex
number, $z^{*}$ denotes its complex-conjugate, and $\bI_N$ is the $N \times N$ identity matrix. Since the imaginary
  part of the principal logarithm is bounded in $(-\pi,\pi]$, the right hand side of eq. (\ref{eod}) is
not extensive and, consequently, unfit to count the number of eigenvalues within $[a,b]$
for single realizations of $\bM$. This issue
comes from our naive derivation of eq. (\ref{eod}), where we assume that the principal complex logarithm fulfills the same
standard properties as those valid for the logarithm of real numbers. In spite of that, this is a necessary step to apply the replica
method. We will see that, after calculating the ensemble average and introducing an
appropriate order-parameter, the problem factorises over sites and the extensivity of the
moments of $I_N(a,b)$ is restored. This procedure is heuristic, but it yields correct results for the moments
of $I_N(a,b)$ for large $N$. Thus, eq. (\ref{eod}) completely encodes the statistics of
 $I_N(a,b)$, even though it is unsuitable to obtain $I_N(a,b)$ for single instances of $\bM$.  

Inserting eq. (\ref{eod}) in eq. (\ref{ssqw}), we find
\begin{equation}
\fl
\mathcal{G}_N (\mu) = \lim_{\eta \rightarrow 0^{+}} \left\langle   \left[ Z(z_b) Z(z_a^{*})  \right]^{\frac{\mu}{2 \pi}}  \left[ Z(z_b^{*}) Z(z_a)  
\right]^{- \frac{\mu}{2 \pi}}     \right\rangle .
\end{equation}
At this point we invoke the replica method in the form
\begin{equation}
\mathcal{G}_N (\mu) = \lim_{\eta \rightarrow 0^{+}} \lim_{n_{\pm} \rightarrow \pm \frac{\mu}{2 \pi} } \mathcal{G}_N (n_{\pm},\eta)   \,,
\end{equation}
in which we have introduced the function
\begin{equation}
\mathcal{G}_N (n_{\pm},\eta)  =
\left\langle   \left[ Z(z_b) Z(z_a^{*})  \right]^{n_{+}}  \left[ Z(z_b^{*}) Z(z_a)  \right]^{n_{-}}     \right\rangle 
\label{jsla}
\end{equation}
for finite $\eta$.
The idea consists in assuming that $n_{\pm}$ are positive integers, which allows to calculate $\mathcal{G}_N (n_{\pm},\eta)$ for $N \gg 1$ 
through a saddle-point
integration. After we have derived the behaviour of $\mathcal{G}_N (n_{\pm},\eta)$ for $N \gg 1$, we take the replica limit $n_{\pm} \rightarrow \pm \frac{\mu}{2 \pi} $
and reconstruct the original function $\mathcal{G}_N (\mu)$ in the limit $\eta \rightarrow 0^{+}$. This is nothing more than the general strategy of the replica approach \cite{BookParisi}. The
only difference lies in the fact that we continue the arbitrary positive integers $n_{\pm}$ to purely imaginary numbers.


\section{Replica method and finite size corrections} \label{bla1}

Our aim in this section is to calculate  $\mathcal{G}_N (n_{\pm},\eta)$ for large but finite $N$. Firstly, we have
to recast eq. (\ref{jsla}) into an exponential form, which is suitable to perform the ensemble average. This is achieved
by representing the functions $Z(z_{a,b})$ and $Z(z_{a,b}^{*})$ as Gaussian integrals. For instance, the function
$Z(z_{a})$ is written as the multidimensional Gaussian integral \cite{negele}
\begin{equation}
\fl
Z(z_{a}) = \frac{1}{\det{\left(\bM - \bI_N z_a \right)}} = \frac{1}{(2 \pi )^N} \int \left( \prod_{i=1}^N d \phi_i d \phi_i^{*}   \right)
\exp{\left[  - i \sum_{ij=1}^N \phi_{i}^{*} \left( \bM - \bI_N z_a  \right)_{ij} \phi_{j}  \right]},
\end{equation}
where $\phi_1,\dots, \phi_N$ are complex integration variables. This representation of $Z(z_{a})$
is appropriate to deal with both the GOE ensemble and the GUE ensemble in the
same framework. Introducing analogous identities for $Z(z_b)$, $Z(z_b^{*})$ and $Z(z_a^{*})$, eq. (\ref{jsla}) can be compactly written as
\begin{equation}
\fl
\mathcal{G}_N (n_{\pm},\eta)  = \int \left( \prod_{i=1}^N d \bvphi_i d \bvphi_i^{\dagger}   \right)
\exp{\left(   i \sum_{i=1}^{N} \bvphi_{i}^{\dagger}. \bZ \bvphi_{i}   \right)}
\left\langle  \exp{\left[ - i \sum_{ij =1}^{N} M_{ij}  \left( \bvphi_{i}^{\dagger}. \bA \bvphi_{j} \right)   \right] }    \right\rangle,
\label{ttpo}
\end{equation}
where the following $2(n_+ + n_{-}) \times 2(n_+ + n_{-})$ block matrices have been introduced
\[ 
\fl
\bA = \left( \begin{array}{cccc}
\bI_{+} & 0 & 0 & 0 \\
0 & \bI_{-}  & 0 & 0 \\
0 & 0 & -\bI_{+} & 0 \\
0 & 0 & 0 & -\bI_{-}
\end{array} \right),
\qquad
\bZ = \left( \begin{array}{cccc}
z_b \bI_{+} & 0 & 0 & 0 \\
0 & z_a \bI_{-}  & 0 & 0 \\
0 & 0 & - z_a^{*} \bI_{+} & 0 \\
0 & 0 & 0 & - z_b^{*}\bI_{-}
\end{array} \right),
\]
with $\bI_{+}$ ($\bI_{-}$) denoting the $n_+ \times n_+$ ($n_- \times n_-$) identity matrix. The integration variables in
eq. (\ref{ttpo}) are $2(n_+ + n_{-})$-dimensional complex vectors in the replica space, defined according to
\[ 
 \bvphi_{i} = \left( \begin{array}{c}
\bphi_{1,i} \\
\bpsi_{1,i}  \\
\bphi_{2,i}  \\
\bpsi_{2,i}
\end{array} \right)
\qquad
i=1,\dots,N ,
\]
where $\bphi_{1,i}$ and $\bphi_{2,i}$ have dimension $n_+$, while $\bpsi_{1,i}$ and $\bpsi_{2,i}$ have dimension $n_-$. Each one of the 
vectors $(\bphi_{1,i},\bphi_{2,i},\bpsi_{1,i}, \bpsi_{2,i})$ 
comes from the Gaussian integral representation of a single function $Z(\dots)$ in eq. (\ref{jsla}). The off-diagonal blocks
of $\bZ$ and $\bA$, filled with zeros, have suitable dimensions, such that their product with $\bvphi_{i}$ is well-defined. 

The ensemble average in eq. (\ref{ttpo}) is easily performed for the GOE and the GUE ensembles by using eqs. (\ref{goe}) and (\ref{gue}), leading to
\begin{equation}
\fl
\mathcal{G}_N (n_{\pm},\eta)  = \int \left( \prod_{i=1}^N d \bvphi_i d \bvphi_i^{\dagger}   \right)
\exp{\left[   i \sum_{i=1}^{N} \bvphi_{i}^{\dagger}. \bZ \bvphi_{i}   - \frac{1}{2N} \sum_{ij =1}^{N} K_{\beta}(\bvphi_{i}, \bvphi_{i}^{\dagger},\bvphi_{j}, \bvphi_{j}^{\dagger})   \right] }  \,,
\label{hhqpr}
\end{equation}  
with the kernel $K_{\beta}$ 
\begin{equation}
\fl
K_{\beta}(\bvphi_{i}, \bvphi_{i}^{\dagger},\bvphi_{j}, \bvphi_{j}^{\dagger}) = |\bvphi_{i}^{\dagger}. \bA \bvphi_{j}|^2 + (2-\beta) \, {\rm Re} \, (\bvphi_{i}^{\dagger}. \bA \bvphi_{j})^2\,,
\label{kernel}
\end{equation} 
which depends on the random matrix ensemble through the  Dyson index $\beta$: $\beta=1$ for the GOE ensemble and $\beta=2$ for the GUE ensemble.
Following the standard approach to decouple the sites in eq. (\ref{hhqpr}), we introduce the order-parameter 
\begin{equation}
\rho(\bvphi,\bvphi^{\dagger}) = \frac{1}{N} \sum_{i=1}^{N} \delta(\bvphi - \bvphi_i   ) \delta(\bvphi^{\dagger} - \bvphi_i^{\dagger}  )
\end{equation}
through a functional Dirac delta, yielding
\begin{eqnarray}
\fl
\mathcal{G}_N (n_{\pm},\eta)  &=& \int \mathcal{D}\rho \mathcal{D}\hat{\rho} \exp{\left[i \int d \bvphi d \bvphi^{\dagger}  \rho(\bvphi,\bvphi^{\dagger})  \hat{\rho}(\bvphi,\bvphi^{\dagger}) \right]} 
\nonumber \\
\fl
& \times& \exp{\left[ - \frac{N}{2}  \int d \bvphi_1 d \bvphi_1^{\dagger} d \bvphi_2 d \bvphi_2^{\dagger} \,  \rho(\bvphi_1,\bvphi_1^{\dagger})    
K_{\beta}(\bvphi_{1}, \bvphi_{1}^{\dagger},\bvphi_{2}, \bvphi_{2}^{\dagger})  \rho(\bvphi_2,\bvphi_2^{\dagger})   \right]} 
\nonumber \\
\fl
& \times& \exp{\left[
N \ln{\left[ \int d \bvphi d \bvphi^{\dagger}  \exp{\left(i \bvphi^{\dagger}. \bZ \bvphi - \frac{i}{N} \hat{\rho}(\bvphi,\bvphi^{\dagger}) \right) }        \right]} \right]} ,
\end{eqnarray}
where $\mathcal{D}\rho \mathcal{D}\hat{\rho}$ is the functional integration measure over $\rho$ and its conjugate order-parameter $\hat{\rho}$. After
performing the Gaussian integral over $\rho$ and introducing a new integration variable $\Phi$
\begin{equation}
\hat{\rho}(\bvphi,\bvphi^{\dagger}) = -i N \int d \bvphi_1 d \bvphi_1^{\dagger} K_{\beta}(\bvphi, \bvphi^{\dagger},\bvphi_{1}, \bvphi_{1}^{\dagger})
\Phi(\bvphi_1,\bvphi_1^{\dagger}) \,,
\end{equation}
we obtain the compact expression 
\begin{equation}
\mathcal{G}_N (n_{\pm},\eta) = \sqrt{\det{K_\beta}} \int \mathcal{D}\Phi \exp{\left(N S_N[\Phi]   \right)} \,,
\label{jaqpm}
\end{equation}
with the action:
\begin{eqnarray}
\fl
S_N[\Phi] &=& \frac{1}{2} \int d \bvphi_1 d \bvphi_1^{\dagger} d \bvphi_2 d \bvphi_2^{\dagger} \,  \Phi(\bvphi_1,\bvphi_1^{\dagger})    
K_{\beta}(\bvphi_{1}, \bvphi_{1}^{\dagger},\bvphi_{2}, \bvphi_{2}^{\dagger})  \Phi(\bvphi_2,\bvphi_2^{\dagger}) \nonumber \\
\fl
&+& \ln{\left[ \int d \bvphi d \bvphi^{\dagger}  \exp{\left(i \bvphi^{\dagger}. \bZ \bvphi -  \int d \bvphi_1 d \bvphi_1^{\dagger} K_{\beta}(\bvphi, \bvphi^{\dagger},\bvphi_{1}, \bvphi_{1}^{\dagger})
\Phi(\bvphi_1,\bvphi_1^{\dagger})  \right) }        \right]}\,.
\label{action}
\end{eqnarray}
The functional integration measure in eq. (\ref{jaqpm}) can be intuitively written as 
\begin{equation}
\mathcal{D}\Phi = \prod_{\bvphi \bvphi^{\dagger}} (- i) \sqrt{\frac{N}{2 \pi}} d \Phi(\bvphi,\bvphi^{\dagger})\,,
\end{equation}
in which the product runs over all possible arguments of the function $\Phi(\bvphi,\bvphi^{\dagger})$.

The next step consists in solving the integral in eq. (\ref{jaqpm}) using the saddle-point method. In the limit $N \rightarrow \infty$, this integral
is dominated by the value $\Phi_0$ that extremizes the action $S_N[\Phi]$, i.e.
\begin{eqnarray}
\fl
\frac{\delta S_N[\Phi]}{\delta \Phi( \bvphi, \bvphi^{\dagger}) } \Bigg{|}_{\Phi=\Phi_0} = 0 , \nonumber \\
\fl
\Phi_0 ( \bvphi, \bvphi^{\dagger}) = \frac{\exp{\left[ i \bvphi^{\dagger}. \bZ  \bvphi - \int  d \bvphi_1 d \bvphi_1^{\dagger}  K_{\beta}(\bvphi, \bvphi^{\dagger},\bvphi_{1}, \bvphi_{1}^{\dagger})
\Phi_0 ( \bvphi_{1}, \bvphi_{1}^{\dagger})\right]  }}{   
\int d \bvphi_2 d \bvphi_2^{\dagger}
\exp{ \left[ i \bvphi_{2}^{\dagger}. \bZ  \bvphi_{2} - \int  d \bvphi_1 d \bvphi_1^{\dagger}  K_{\beta}(\bvphi_{2}, \bvphi_{2}^{\dagger},\bvphi_{1}, \bvphi_{1}^{\dagger})
\Phi_0 ( \bvphi_{1}, \bvphi_{1}^{\dagger}) \right]  }}\,.
\label{hdqpkk}
\end{eqnarray}
We are not interested only on the behaviour of $\mathcal{G}_N$ strictly in the limit $N \rightarrow \infty$, but also on the first 
perturbative correction due to large
but finite $N$. We will see that such correction yields precisely the correct logarithmic scaling of the variance of $I_N(a,b)$ with 
respect to $N$.
It is important to note that, up to now, we have not done any approximation regarding the system
size dependence. In other words, eqs. (\ref{jaqpm}) and (\ref{action}) are exact for finite $N$. As we can see from eq. (\ref{action}), the only source 
of finite size  fluctuations in the action comes from fluctuations of the order-parameter around its $N \rightarrow \infty$ 
limit $\Phi_0$. Let us assume that such fluctuations are of the form \cite{MetzPar}
\begin{equation}
\Phi ( \bvphi, \bvphi^{\dagger}) = \Phi_0 ( \bvphi, \bvphi^{\dagger}) - \frac{1}{\sqrt{N}} \chi(\bvphi, \bvphi^{\dagger}).
\label{rem}
\end{equation}
After expanding $S_N[\Phi]$ around $\Phi_0$ up to $O(1/N)$, we substitute eq. (\ref{rem}) and rewrite eq. (\ref{jaqpm}) as follows
\begin{eqnarray}
\fl
\mathcal{G}_N (n_{\pm},\eta) = \sqrt{\det{K_\beta}} \exp{\left( N S_0[\Phi_0] \right)} \nonumber \\
\fl
\times
\int \mathcal{D}\chi \exp{\left[
\frac{1}{2} \int d \bvphi_{1} d \bvphi_{1}^{\dagger}  d \bvphi_{2} d \bvphi_{2}^{\dagger} \chi(\bvphi_{1} , \bvphi_{1}^{\dagger})  H(\bvphi_{1} , \bvphi_{1}^{\dagger}, \bvphi_{2} , \bvphi_{2}^{\dagger} )
\chi(\bvphi_{2} , \bvphi_{2}^{\dagger}) 
\right]} \,,
\label{hgqpo}
\end{eqnarray}
in which 
\begin{eqnarray}
H(\bvphi_{1} , \bvphi_{1}^{\dagger}, \bvphi_{2} , \bvphi_{2}^{\dagger} ) = 
\frac{\delta^2 S_N }{\delta \Phi( \bvphi_1, \bvphi_{1}^{\dagger}) \delta \Phi( \bvphi_2, \bvphi_{2}^{\dagger})  } \Bigg{|}_{\Phi=\Phi_0}\,.
\label{yuyr}
\end{eqnarray}
The leading contribution to the action, defined as $S_0[\Phi_0]$, is formally given by eq. (\ref{action}) with $\Phi$ replaced by $\Phi_0$, while the functional integration measure
in eq. (\ref{hgqpo}) reads 
\begin{equation}
\mathcal{D}\chi = \prod_{ \bvphi, \bvphi^{\dagger}} i \sqrt{\frac{1}{2 \pi}} d \chi(\bvphi, \bvphi^{\dagger} ) \,.
\end{equation}
By computing explicitly the derivatives in eq. (\ref{yuyr}) and using eq. (\ref{hdqpkk}), the matrix $\bH$ of second
derivatives can be expressed as
\begin{equation}
\bH = \bK_{\beta} + \bK_{\beta} \bT \,,
\label{tuyyn}
\end{equation}
with the elements of $\bT$ defined according to
\begin{eqnarray}
\fl
T(\bvphi_{1} , \bvphi_{1}^{\dagger},\bvphi_{2} , \bvphi_{2}^{\dagger}  ) &=&
K_{\beta} (\bvphi_{1} , \bvphi_{1}^{\dagger},\bvphi_{2} , \bvphi_{2}^{\dagger}) \Phi_0 (\bvphi_{1} , \bvphi_{1}^{\dagger} ) \nonumber \\
\fl
&-& \Phi_0 (\bvphi_{1} , \bvphi_{1}^{\dagger} ) \int d \bvphi d \bvphi^{\dagger} K_{\beta} (\bvphi_{2} , \bvphi_{2}^{\dagger},\bvphi , \bvphi^{\dagger}) \Phi_0 (\bvphi , \bvphi^{\dagger} )\,.
\label{hapq}
\end{eqnarray}
Now we are ready to obtain an expression for $\mathcal{G}_N$ when $N$ is large but finite. After integrating over the Gaussian fluctuations in eq. (\ref{hgqpo}), we
substitute eq. (\ref{tuyyn}) and derive
\begin{equation}
\mathcal{G}_N (n_{\pm},\eta) = \exp{\left( N S_0[\Phi_0]  + \frac{1}{2} \sum_{\ell=1}^{\infty} \frac{(-1)^{\ell}}{\ell}  {\rm Tr} \, \bT^{\ell}  \right)}\,.
\label{uya}
\end{equation}
The above equation is determined essentially by the behaviour of $\Phi_0$, which is obtained from the solution of eq. (\ref{hdqpkk}). In the next section we
show the outcome for $\mathcal{G}_N$ when $\Phi_0$ is symmetric with respect to the interchange of the replica indexes.

\section{The replica symmetric characteristic function} \label{bla2}

The next step in our calculation of the characteristic function consists in performing the replica limit $n_{\pm} \rightarrow \pm \mu/2\pi$ of eq. (\ref{uya}). Thus, we need 
to understand how $\mathcal{G}_N (n_{\pm},\eta)$ depends on $n_{\pm}$, which is ultimately determined by the solutions of eq. (\ref{hdqpkk}). In order
to proceed further, we follow previous works \cite{Kuhn2008,MetzPar,Metz2015,MetzLett} and we make the following Gaussian {\it ansatz} for the order-parameter
\begin{equation}
\Phi_0(\bvphi , \bvphi^{\dagger}) = \frac{\det{\bC}}{(2 \pi i)^{2 (n_{+}+ n_{-}) } } \exp{\left( - \bvphi^{\dagger} . \bC \bvphi  \right)} \,,
\label{RS}
\end{equation}
which is parametrised by the $2(n_{+} + n_{-}) \times 2(n_{+} + n_{-})$ diagonal block matrix 
\[ 
\fl
\bC = \left( \begin{array}{cccc}
\bI_{+} \Delta_{b}^{*} & 0 & 0 & 0 \\
0 & \bI_{-} \Delta_{a}^{*} & 0 & 0 \\
0 & 0 & \bI_{+} \Delta_{a} & 0 \\
0 & 0 & 0 & \bI_{-} \Delta_{b}
\end{array} \right), \qquad {\rm Re} \Delta_{a} > 0 \qquad {\rm Re} \Delta_{b} > 0 \,,
\]
given in terms of the complex parameters $\Delta_a$ and $\Delta_b$. The conditions 
${\rm Re} \Delta_{a} > 0$ and ${\rm Re} \Delta_{b} > 0$  ensure the convergence of any  Gaussian integrals over $\Phi_0$.
The off-diagonal blocks in $\bC$ have suitable dimensions such that the standard matrix operations involving $\bC$ and $\bvphi$ are well-defined. The above
assumption for $\Phi_0$ is symmetric with respect to the permutation of replicas inside each group $1 , \dots, n_{+}$ and $ 1 , \dots, n_{-}$. We do not
consider here solutions of eq. (\ref{hdqpkk}) that break replica symmetry.

Let us explore the consequences of the replica symmetric (RS) assumption. Substituting eq. (\ref{RS}) in eq. (\ref{hdqpkk}), considering the
explicit form of the kernel $K_{\beta}$ (see eq. (\ref{kernel})), and noting that
\begin{eqnarray}
\fl
\int  d \bvphi_1  d \bvphi_{1}^{\dagger} \, | \bvphi^{\dagger}. \bA \bvphi_{1} |^2 \, \Phi_0(\bvphi_1 , \bvphi_{1}^{\dagger}) =  \bvphi^{\dagger}. \bC^{-1} \bvphi \,, \nonumber \\
\fl
\int  d \bvphi_1  d \bvphi_{1}^{\dagger} \, {\rm Re} ( \bvphi^{\dagger}. \bA \bvphi_{1} )^2 \, \Phi_0(\bvphi_1 , \bvphi_{1}^{\dagger}) = 0\,,
\end{eqnarray}
we conclude that the RS form of $\Phi_0$ solves the self-consistent eq. (\ref{hdqpkk}) provided the parameters
$\Delta_{a}$ and $\Delta_{b}$ fulfill the quadratic equations
\begin{equation}
\Delta_{a}^2 - i z_{a}^{*}  \Delta_{a} - 1 = 0\,, \qquad \Delta_{b}^2 - i z_{b}^{*}  \Delta_{b} - 1 = 0\,.
\label{quadr}
\end{equation}

Now we are in a position to derive the explicit dependency of $\mathcal{G}_N (n_{\pm},\eta)$ with respect to $n_{\pm}$. Inserting the RS
assumption for $\Phi_0$ in eq. (\ref{action}), the leading contribution to the action is derived
\begin{equation}
\fl
S_0(n_{\pm}) = \frac{1}{2} \left[ \frac{n_{+}}{\left(\Delta_{b}^{*}\right)^2} + \frac{n_{-}}{\left(\Delta_{a}^{*}\right)^2} 
+ \frac{n_{+}}{\Delta_{a}^2} + \frac{n_{-}}{\Delta_{b}^2}  \right] - n_+ \ln{\left( \Delta_a \Delta_{b}^*  \right)} - n_- \ln{\left( \Delta_a^{*} \Delta_{b}  \right)}\,. 
\label{ghpm}
\end{equation}
The second contribution appearing in eq. (\ref{uya}) involves an infinite series, so that we have to evaluate the RS form of
the coefficients ${\rm Tr} \bT^{\ell}$. Plugging eq. (\ref{RS}) in eq. (\ref{hapq}) and performing the Gaussian integrals, we have derived the following expression
\begin{eqnarray}
\fl
{\rm Tr} \bT^{\ell} &=& (2 - \beta) \left[ \frac{n_{+}}{\left(\Delta_{b}^{*}\right)^{2 \ell}} + \frac{n_{-}}{\left(\Delta_{a}^{*}\right)^{2 \ell}} 
+ \frac{n_{+}}{\Delta_{a}^{2 \ell}} + \frac{n_{-}}{\Delta_{b}^{2 \ell}}  \right]  \nonumber \\
\fl
&+&
\frac{2}{\beta} \left[ \frac{n_{+}}{\left(\Delta_{b}^{*}\right)^{\ell}} + \frac{n_{-}}{\left(\Delta_{a}^{*}\right)^{\ell}} 
+ \frac{(-1)^{\ell} n_{+}}{\Delta_{a}^{\ell}} + \frac{(-1)^{\ell} n_{-}}{\Delta_{b}^{\ell}}  \right]^2\,,
\label{jjqp}
\end{eqnarray}
in which we have used the fact that the Dyson index is limited to the values $\beta=1$ or $\beta=2$. Finally, eqs. (\ref{ghpm}) and (\ref{jjqp}) are substituted in eq. (\ref{uya}), the 
limit $n_{\pm} \rightarrow \pm \mu/2 \pi$ is taken, and the following expression for the characteristic function is
obtained
\begin{equation}
\mathcal{G}_N (\mu) = \lim_{\eta \rightarrow 0^{+}} \exp{\left[  i \mu \left \langle I_N \right  \rangle_{\eta}    
- \frac{\mu^2}{2} \left(  \left \langle  I_N^2  \right \rangle_{\eta}  -  \left \langle I_N  \right \rangle_{\eta}^{2}   \right) \right]} \,,
\end{equation}
where $\left \langle  I_N  \right \rangle_{\eta}$ and $\left \langle  I_N^2  \right \rangle_{\eta}  -  \left \langle I_N  \right \rangle_{\eta}^{2} $ are, respectively, the
mean and the variance of $I_N(a,b)$ for finite $\eta > 0$
\begin{eqnarray}
\fl
\frac{\left \langle  I_N  \right \rangle_{\eta}}{N}  &=& \frac{i}{4 \pi} \left[ \frac{1}{\Delta_b^2} - \frac{1}{(\Delta_b^{*})^2} - \frac{1}{\Delta_a^2} + \frac{1}{(\Delta_a^{*})^2}  \right]
- \frac{i}{ 2 \pi} \ln{\left( \frac{\Delta_b \Delta_a^{*}  }{ \Delta_b^{*} \Delta_a  }  \right)}  \label{mean} \\
\fl
&-& \frac{i (2 - \beta) }{4 \pi N} \sum_{\ell=1}^{\infty} \frac{(-1)^{\ell}}{\ell} \left[  \frac{1}{\Delta_a^{2 \ell}} - \frac{1}{(\Delta_a^{*})^{2 \ell}} 
-   \frac{1}{\Delta_b^{2 \ell}} +   \frac{1}{(\Delta_b^{*})^{2 \ell}}   \right]\,, \nonumber \\
\fl
 \left \langle  I_N^2  \right \rangle_{\eta}  &-&  \left \langle I_N  \right \rangle_{\eta}^{2} = - \frac{1}{2 \beta \pi^2} \sum_{\ell=1}^{\infty} 
\frac{(-1)^{\ell}}{\ell} \left[   \frac{(-1)^{\ell}}{ \Delta_a^{\ell}}  - \frac{1}{(\Delta_a^{*})^{\ell}} -   \frac{(-1)^{\ell}}{ \Delta_b^{\ell}} 
+  \frac{1}{(\Delta_b^{*})^{\ell}}   \right]^{2} .
\label{var}
\end{eqnarray}
Note that the contribution of $O(N)$ in the exponent of eq. (\ref{uya}) depends linearly on $\mu$, only providing the
mean value $\left \langle I_N \right \rangle$. If one wants to extract any information about the typical fluctuations around $\left \langle I_N \right \rangle$, one has to take
into account the finite size fluctuations of the order-parameter $\Phi$. This is in contrast with some models of sparse random matrices, where the calculation of the leading 
term of the action is enough to obtain the linear scaling of the variance of $I_N$ with $N$ \cite{Metz2015}. As we will see below, here the system size dependence 
of $\lim_{ \eta \rightarrow 0^+} \left \langle  I_N^2  \right \rangle_{\eta}  -  \left \langle I_N  \right \rangle_{\eta}^{2}$ manifests itself as the divergence of
the infinite series present in eq. (\ref{var}). The finite-size fluctuations of $\Phi$ also yield the $O(1/N)$ correction to $\left \langle I_N \right \rangle/N$
appearing in eq. (\ref{mean}).

\section{The mean and the variance of $I_N$} \label{bla3}

In this section we derive explicit analytical results from eqs. (\ref{mean}) and (\ref{var}). The solutions of eqs. (\ref{quadr}) read
\begin{equation}
\Delta_a = \frac{1}{2} \left( i z_{a}^{*} \pm \sqrt{4 - (z_{a}^{*})^2 }    \right),  \qquad \Delta_b = \frac{1}{2} \left( i z_{b}^{*} \pm \sqrt{4 - (z_{b}^{*})^2 }    \right).
\end{equation}
We consider below the behaviour of $\Delta_a$ and $\Delta_b$ in the limit $\eta \rightarrow 0^+$ for specific observables 
depending on the values of $a$ and $b$.
For this purpose, it is instrumental to recognise that the 
eigenvalues of the GOE and the GUE random matrix ensembles are distributed, for $N \rightarrow \infty$, according 
to the Wigner semicircle law with support in $[-2,2]$ \cite{BookMehta}.

\subsection{The index}

The first observable considered here is the index, i.e., the number of eigenvalues 
smaller than a certain threshold $-2 < b < 2$. In this
case we set $a < -2$ and $0 \leq |b| < 2$, which leads to the following solutions for $\eta \rightarrow 0^{+}$ 
\begin{eqnarray}
\Delta_a &=& |\Delta_a| e^{-i \frac{\pi}{2}}  \label{ssq},  \\
\Delta_b &=& e^{i \theta_b} ,
\label{ddpq}
\end{eqnarray}
with the argument: 
\begin{equation}
\theta_b = {\rm arctan} \left(  \frac{b}{\sqrt{4 - b^2} } \right) .
\end{equation}
After plugging eqs. (\ref{ssq}) and (\ref{ddpq}) in eq. (\ref{mean}), we can sum the convergent series and obtain
\begin{equation}
\frac{\left \langle  I_N  \right \rangle}{N}  = \frac{1}{2} + \frac{1}{2 \pi} \sin{(2 \theta_b)} + \frac{1}{\pi} \theta_b 
+ \frac{(2 - \beta)}{2 N} C(\theta_b) ,
\label{ffg}
\end{equation}
where
\begin{equation}
C(\theta) = \frac{i}{2 \pi} \ln{\left( \frac{1 + e^{2 i \theta}  }{1 + e^{-2 i \theta} }  \right)}.
\label{sqwbb}
\end{equation}

\begin{figure}[H]
\center
\includegraphics[scale=1.0]{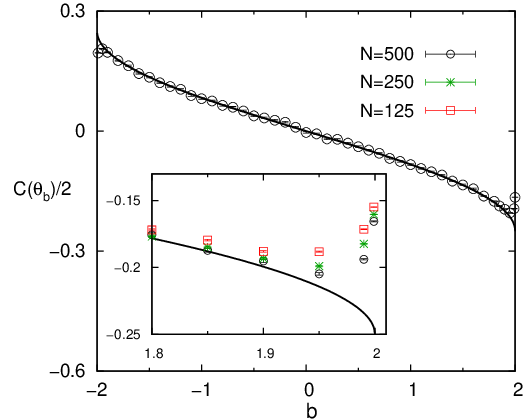}
\caption{Comparison between the analytical result (solid line) for the $O(1/N)$ coefficient of eq. (\ref{ffg}) with numerical diagonalisation
of $N \times N$ random matrices drawn from the GOE ensemble. The inset shows the behaviour for different $N$ when the upper band edge is approached.
}
\label{correcindex}
\end{figure}

The leading term of eq. (\ref{ffg}) agrees with an exact  result \cite{Perez2014b}.
The coefficient $C(\theta_b)$ can be 
derived from the integral $\int_{-\infty}^{b }d \lambda  \, \rho_{1}(\lambda)$, where $\rho_{1}(\lambda)$ is
the $O(1/N)$ correction to the average spectral density. In the case of $\beta=1$ (GOE), eq. (\ref{ffg})
is in full agreement with references \cite{verbaar1984,dhesi1990,Itoi1997,Kalisch2002}, where 
$\rho_{1}(\lambda)$ is computed exactly using replicas \cite{verbaar1984,dhesi1990,Itoi1997} and supersymmetry \cite{Kalisch2002}.
In figure \ref{correcindex} we also compare the analytical result for $C(\theta_b)$ with numerical
diagonalisation results for real symmetric matrices drawn from the GOE ensemble. The agreement between our analytical expression and numerical
diagonalisation is very good until the band edge $b=2$ is approached, where finite size fluctuations become stronger and a discrepancy between 
theory and numerics is evident. The results shown on the inset exhibit the convergence of the numerical
results to the analytical formula as $N$ increases.

For $\beta=2$ (GUE), the $O(1/N)$ correction in eq. (\ref{ffg}) is zero. This 
is in contrast to the exact expression for $\rho_{1}(\lambda)$ obtained through supersymmetry \cite{Kalisch2002}, replicas \cite{Kamenev} and
the large $N$ expansion of orthogonal polynomials \cite{Kamenev}. These methods
show that $\rho_{1}(\lambda)$ is an oscillatory function of $\lambda$ with $N$ maxima. For large but finite  $N$, the 
integration of  $\rho_{1}(\lambda)$ over an interval within the support of the spectral density yields 
a negligible contribution to the $O(1/N)$ correction of $\frac{\left \langle  I_N  \right \rangle}{N}$.
This has been confirmed 
by computing $\left \langle  I_N  \right \rangle / N$  through numerical diagonalisation and then
subtracting the leading term of eq. (\ref{ffg}), which yields an outcome orders of magnitude smaller than $1/N$.
The present approach is not able to recover the correct $O(1/N)$ contribution
to $\left \langle  I_N  \right \rangle / N$ due to our replica symmetric assumption for the
order parameter. This is evident  from the calculation of the spectral density using
the fermionic replica method \cite{Kamenev}, in which the oscillatory correction $\rho_1(\lambda)$ is derived by including saddle-points
that break replica symmetry.

Let us derive an expression for the variance of the index.  Inserting the above expressions for $\Delta_a$ and $\Delta_b$ in eq. (\ref{var}) and performing
the sum of the convergent series, we can write the formal expression
\begin{equation}
\left \langle  I_N^2  \right \rangle  -  \left \langle I_N  \right \rangle^{2} = \frac{1}{\beta \pi^2} 
\left[ \sum_{\ell=1}^{\infty} \frac{1}{\ell} + \frac{1}{2} \ln{\left( 2 + 2 \cos{(2 \theta_b)}   \right)}    \right]\,.
\label{haq}
\end{equation}
We have isolated in eq. (\ref{haq}) the divergent contribution to the variance in the form of the harmonic series. This is
not surprising, since the leading term of $\left \langle  I_N^2  \right \rangle  -  \left \langle I_N  \right \rangle^{2}$ 
should indeed diverge for $N \rightarrow \infty$. The question here is how $\left \langle  I_N^2  \right \rangle  -  \left \langle I_N  \right \rangle^{2}$  scales with $N$.
In order to extract this behaviour in the replica framework, the authors of \cite{Cavagna2000} keep the regularizer $\eta$ finite until the end of 
the calculation, and then assume there is a 
functional relation between $\eta$ and $N$. Here a different strategy has been pursued, where the limit $\eta \rightarrow 0^{+}$ has 
been performed before considering the convergence of  the series in eqs. (\ref{mean}) and (\ref{var}).
This approach gives rise to the divergent contribution in eq. (\ref{haq}), which is naturally interpreted as the leading 
term $\lim_{N \rightarrow \infty} \left \langle  I_N^2  \right \rangle  -  \left \langle I_N  \right \rangle^{2}$.
Thus, in order to understand how the variance scales with $N$, we have to study how the harmonic series diverges. For large $N$, the partial
summation behaves as
\begin{equation}
\sum_{\ell=1}^{N} \frac{1}{\ell} = \ln N + \gamma + O(1/N) \,,
\end{equation}
where $\gamma$ is the Euler-Mascheroni constant. Consequently, we conclude that the variance behaves for $N \gg 1$ as follows
\begin{equation}
\left \langle  I_N^2  \right \rangle  -  \left \langle I_N  \right \rangle^{2} = \frac{1}{\beta \pi^2} 
\left[ \ln N  + \gamma + \frac{1}{2} \ln{\left( 2 + 2 \cos{(2 \theta_b)}   \right)}    \right]\,.
\label{hhp}
\end{equation}
The above equation recovers exactly the leading behaviour of the index variance for $N \gg 1$ \cite{Majumdar1,Majumdar2,Isaac2016,Perez2014b}. However, eq. (\ref{hhp}) fails
in reproducing the $O(1)$ correction to $\left \langle  I_N^2  \right \rangle  -  \left \langle I_N  \right \rangle^{2}$. For $b=0$ and $\beta=2$, the $O(1)$ term
in eq. (\ref{hhp}) is given by $(\gamma + \ln 2)/2 \pi^2$, which is only part of the exact result for the GUE ensemble \cite{Majumdar2}. For $\beta=1$, the $O(1)$ term of 
eq. (\ref{hhp}) does not agree as well with the available result of \cite{Cavagna2000}, obtained from a fitting of numerical diagonalisation results. 
As we shall discuss below, this inaccuracy comes from the replica-symmetric assumption for the order-parameter. 

\subsection{The number of eigenvalues in a symmetric interval}

For the second observable we set $a = -L$ and $b=L$, with $0 < L < 2$. In this case the random variable $I_N$ 
quantifies the number of eigenvalues within $[-L,L]$. The solutions for $\Delta_a$ and $\Delta_b$ 
in the limit $\eta \rightarrow 0^+$ read
\begin{eqnarray}
\Delta_a = \Delta_{b}^{*}, \label{d1} \\
\Delta_b = e^{i \theta_L}, \quad \theta_L = {\rm arctan} \left(  \frac{L}{\sqrt{4 - L^2  } }  \right).
\label{d2}
\end{eqnarray}
Inserting the above forms in eq. (\ref{mean}) and summing the series we obtain
\begin{equation}
\frac{\left \langle  I_N  \right \rangle}{N}  =
\frac{1}{\pi} \sin{(2 \theta_L )} + \frac{2}{\pi} \theta_L + \frac{(2-\beta)}{N } C(\theta_L),
\label{ssqa}
\end{equation}
with $C(\theta_L)$ defined in eq. (\ref{sqwbb}).

\begin{figure}[h]
\center
\includegraphics[scale=1.0]{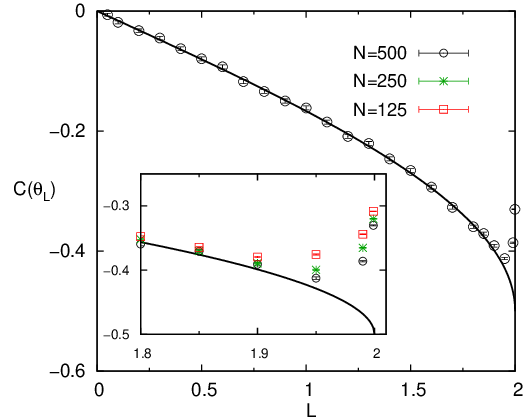}
\caption{Comparison between the analytical result (solid line) for the $O(1/N)$ term of eq. (\ref{ssqa}) with numerical diagonalisation
of $N \times N$ random matrices drawn from the GOE ensemble. The inset shows the behaviour for different $N$ when the upper band edge is approached.
}
\label{correc}
\end{figure}

The leading term $\lim_{N \rightarrow \infty} \frac{\left \langle  I_N  \right \rangle}{N}$ of eq. (\ref{ssqa}) agrees with the exact result \cite{Marino2014}. 
For $\beta =2$, the $O(1/N)$ correction in the above equation is absent, due to the replica
symmetric {\it ansatz} for the order parameter. The $O(1/N)$ contribution to $\frac{\left \langle  I_N  \right \rangle}{N}$
can be derived by integrating the $O(1/N)$ correction to the spectral density over $[-L,L]$. In the replica framework, the latter
quantity is exactly calculated only when replica symmetry breaking is taken into account \cite{Kamenev}, i.e., the 
situation here is completely analogous
to the case of the index discussed above.
For $\beta=1$, our result for the $O(1/N)$ correction in eq. (\ref{ssqa}) can be derived
from references \cite{verbaar1984,dhesi1990,Itoi1997,Kalisch2002}, where the $O(1/N)$
contribution to the spectral density is computed exactly using replicas \cite{verbaar1984,dhesi1990,Itoi1997} and
supersymmetry \cite{Kalisch2002}. In
figure \ref{correc} we compare the analytical behaviour of $C(\theta_L)$ with numerical
diagonalisation of real symmetric matrices drawn from the GOE ensemble. 
Similarly to the $O(1/N)$ correction to the average index, figure  \ref{correc} illustrates the convergence
of the numerical diagonalisation results to the analytical formula of  $C(\theta_L)$ for increasing $N$. 
However, the variance of $I_N$ for Gaussian random matrices displays 
an abrupt change of behaviour as we reach the scaling regime $2-L = O(N^{-2/3})$  \cite{Marino2014,Marino2016}, which indicates 
that our formula for the $O(1/N)$ coefficient might in fact breakdown sufficiently close to $L=2$. A detailed analysis of the behaviour
close to the band edges will not be pursued here.

The variance of the number of eigenvalues $I_N$ within $[-L,L]$ is the so-called number variance \cite{BookMehta}. The
substitution of eqs. (\ref{d1}) and (\ref{d2}) in eq. (\ref{var}) reads
\begin{equation}
\left \langle  I_N^2  \right \rangle  -  \left \langle I_N  \right \rangle^{2} =
\frac{2}{\beta \pi^2} \left[ \sum_{\ell = 1}^{\infty} \frac{1}{\ell} + \frac{1}{2} \ln{\left(  \sin^{2}{(2 \theta_L)}  \right)}      \right],
\end{equation}
where the divergent contribution appears once more as a harmonic series. Following the reasoning
of the previous subsection, we conclude that the number variance for $N \gg 1$ is given by
\begin{equation}
\left \langle  I_N^2  \right \rangle  -  \left \langle I_N  \right \rangle^{2} =
\frac{2}{\beta \pi^2} \left[ \ln N + \gamma + \ln{\left(  \sin{(2 \theta_L)}  \right)}      \right].
\label{ffpq}
\end{equation}
The leading contribution for $N \gg 1$ in the above equation is the same as in previous works \cite{Marino2014,Marino2016}, as long as $L$ is not 
too close to the band edge $L=2$ \cite{Marino2014,Marino2016}. 

Recently, the replica method has been used to compute the number variance for the Anderson model
on a regular random graph \cite{MetzLett,Metz2017}. The variance
$\left \langle  I_N^2  \right \rangle  -  \left \langle I_N  \right \rangle^{2}$, when calculated in the
microscopic scale $L=O(1/N)$, allows to clearly distinguish between 
extended, localised and multifractal eigenfunctions \cite{Altshuler88,Mirlin}. One of the central
arguments in \cite{Metz2017} is that the limit $\lim_{L \rightarrow 0^+} \left \langle  I_N^2  \right \rangle  
-  \left \langle I_N  \right \rangle^{2}$ should give the leading behaviour of the number variance in the 
relevant regime $L=O(1/N)$. Equation (\ref{ffpq}) is strictly valid  for $L=O(1)$, independently 
of $N$, but here we can check explicitly this argument by taking the limit $L \rightarrow 0^+$ 
in eq. (\ref{ffpq}) and then comparing the outcome with standard random matrix results \cite{BookMehta}. 
In the regime where $L \rightarrow 0^+$ with $L \gg 1/N$, eq. (\ref{ffpq}) becomes
\begin{equation}
\left \langle  I_N^2  \right \rangle  -  \left \langle I_N  \right \rangle^{2} =
\frac{2}{\beta \pi^2} \left( \ln{s} + \gamma  \right),
\label{fflpq}
\end{equation}
where  $s \equiv LN \gg 1$. The leading term of eq. (\ref{fflpq}) is in perfect agreement with the
standard results for the GOE and the GUE ensembles \cite{BookMehta}, which supports
the essential claim of \cite{Metz2017}.   

\subsection{The number of eigenvalues in an arbitrary interval}

Lastly, let us present analytical results when $|a| < 2$ and $|b| < 2$, with $b > a$. In this situation, $\Delta_a$ and $\Delta_b$ are given by
\begin{eqnarray}
\Delta_a &=& e^{i \theta_a} , \nonumber \\
\Delta_b &=& e^{i \theta_b} , \nonumber 
\end{eqnarray}
where
\begin{equation}
\theta_a = {\rm arctan} \left(  \frac{a}{\sqrt{4 - a^2  } }  \right), \qquad \theta_b = {\rm arctan} \left(  \frac{b}{\sqrt{4 - b^2  } }  \right).
\end{equation}
Inserting the above expressions for $\Delta_a$ and $\Delta_b$ in eqs. (\ref{mean}) and (\ref{var}), we obtain
\begin{eqnarray}
\fl
\frac{\left \langle  I_N  \right \rangle}{N} &=& \frac{1}{2 \pi} \left[  \sin{\left( 2 \theta_b   \right)} - \sin{\left( 2 \theta_a   \right)}   \right]
+ \frac{1}{\pi} \left( \theta_b - \theta_a  \right) + \frac{(2 - \beta)}{2 N} 
\left[ C(\theta_b) - C(\theta_a)  \right], \nonumber \\
\fl
\left \langle  I_N^2  \right \rangle  &-&  \left \langle I_N  \right \rangle^{2} = \frac{2}{\beta \pi^2} \Bigg{\{}
\ln N + \gamma + \frac{1}{4} \ln{\left( 2 + 2 \cos(2 \theta_a )    \right)} + \frac{1}{4} \ln{\left( 2 + 2 \cos(2 \theta_b )    \right)} \nonumber \\
\fl
&-& \frac{1}{2} \ln{\left[ \frac{ 1 + \cos{\left( \theta_a + \theta_b   \right)}  }{1 - \cos{\left( \theta_a - \theta_b   \right)}  }    \right]}
\Bigg{\}},
\label{ffrw}
\end{eqnarray}
in which we have followed a similar calculation as in the previous subsections. We conclude that the leading behaviour of
$\left \langle  I_N^2  \right \rangle  -  \left \langle I_N  \right \rangle^{2}$ for $N \gg 1$ is independent of the interval $[a,b]$.
By setting $a = \lambda - L$ and $b = \lambda + L$ in eq. (\ref{ffrw}), with arbitrary $|\lambda| < 2$, the leading term
of eq. (\ref{ffrw}) converges, in the limit $L \rightarrow 0^{+}$, to the leading contribution of eq. (\ref{fflpq}).
Therefore, eq. (\ref{fflpq}) is not restricted to an interval of size $O(1/N)$ 
around the center of the band $\lambda=0$, but it holds for any $-2 < \lambda < 2$, provided $\lambda$ is not too close to one of the band edges.

\section{Final remarks} \label{bla4}

In this work we have applied the replica approach 
to derive analytical expressions for the
mean and the variance of the number $I_N$ of eigenvalues within a certain interval of the real
line in the case of $N \times N$ Gaussian random matrices.
Although the present method has been discussed in previous works for sparse random
matrices \cite{Metz2015,MetzLett}, here we go one step further and explain how the  fluctuations
of $I_N$ for Gaussian random matrices are recovered if one takes into account
the $O(1/\sqrt{N})$ correction to the order-parameter around its $N \rightarrow \infty$ limit. 
Thus, in this work we have not derived novel analytical results, but we
have carefully assessed the exactness of the replica-symmetric method 
by considering random-matrix ensembles for which many exact analytical results are available.

The universal logarithmic scaling $\langle I_N^2 \rangle - \langle I_N \rangle^2 = O(\ln N)$ ($N \gg 1$)  
of Gaussian random matrices is naturally recovered by studying how the harmonic series diverges. In the limit $L \rightarrow 0^+$, 
eq. (\ref{ffpq}) for the number variance converges to standard results in the regime $L =O(1/N)$ \cite{BookMehta}, i.e., when the size
of the interval $[-L,L]$ is measured in units of the mean level spacing. This strongly suggests that the present method can be used
to inspect the spectral fluctuations at a local scale and, consequently, study the ergodicity of the eigenfunctions, a point raised
in a previous work \cite{Metz2017}. 
 
The reason why our method does not reproduce exactly the $O(1)$ term of $\langle I_N^2 \rangle - \langle I_N \rangle^2$
lies in the replica-symmetric form of the order-parameter.
The variance of $I_N$ is directly related to the two-point
correlation function $R(\lambda,\lambda^{\prime}) = \langle  \rho_N(\lambda)  \rho_N(\lambda^{\prime}) \rangle$ \cite{Mirlin}, with
$\rho_N(\lambda) = N^{-1} \sum_{\alpha=1}^{N} \delta(\lambda - \lambda_\alpha)$. In the fermionic replica method \cite{Kamenev}, the
replica-symmetric saddle-point yields $R(\lambda,\lambda^{\prime}) \propto \left[ N \left( \lambda - \lambda^{\prime}  
\right)\right]^{-2}$ in the regime $|\lambda - \lambda^{\prime}| = O(1/N)$, which gives rise to the leading term
of $\langle I_N^2 \rangle - \langle I_N \rangle^2$.
However, the correct $O(1)$ contribution 
to $\langle I_N^2 \rangle - \langle I_N \rangle^2$ is only obtained
when one includes the oscillatory part of $R(\lambda,\lambda^{\prime})$ \cite{BookMehta}, which has been
exactly derived using both orthogonal polynomials \cite{BookMehta} and the supersymmetric 
approach \cite{Efetov}. In the fermionic replica method, this oscillatory
contribution is obtained only by taking into account saddle-points that break replica-symmetry \cite{Kamenev}.

The present approach also yields an analytical expression for the $O(1/N)$ correction
to the average $\left \langle I_N \right \rangle/N$, which agrees with exact results in the case
of the GOE ensemble \cite{verbaar1984,dhesi1990,Itoi1997,Kalisch2002}.  In the case of the GUE
ensemble, the replica-symmetric saddle-point is not sufficient to recover
the $O(1/N)$ correction to the average $\left \langle I_N \right \rangle/N$. This contribution
arises from the integration of the $O(1/N)$ oscillatory correction to the average 
spectral density, which can be only computed by considering replica symmetry breaking \cite{Kamenev}.

In comparison with the Coulomb gas method \cite{Dyson1}, whose application is limited to rotationally invariant ensembles, the
replica method is more versatile, in the sense it can be applied to  random matrix ensembles where
the joint distribution of eigenvalues is not analytically known. The present paper opens the
door to study the typical eigenvalue fluctuations of a class of random matrix ensembles where, similarly 
to rotationally invariant random matrices, the eigenvalues strongly repel each other, but their joint distribution is not available. 
An important example of this class
of models is the ensemble of random regular graphs \cite{Wormald,Bollobas}, whose fluctuations of $I_N$ we will consider
in a future work. In addition, it would be interesting to 
extend the present formalism  to
include replica symmetry breaking and then compare with the exact results. We leave this matter
for future investigation.

\ack

FLM akcnowledges Isaac P\'erez Castillo for many interesting and important discussions.

\section*{References}
\bibliographystyle{ieeetr} 
\bibliography{biblio}

\end{document}